\documentclass[]{pasj02}
\newif\ifsubmission
\submissionfalse  
\ifsubmission
    \usepackage[switch, mathlines]{lineno} 
\fi
\usepackage{ulem}
\usepackage{calligra}
\usepackage{physics}

\jyear{2024}
\Received{}
\Accepted{}

\newcommand{\msun}{\mathrm{M}_\odot}
\newcommand{\change}[1]{#1}

\newcommand{\add}[1]{#1}
\ifsubmission
\newcommand{\alttxt}[1]{Alt text: #1}
\else
\newcommand{\alttxt}[1]{}
\fi

\begin{document}
  \title{Modeling supernova feedback in galaxy formation simulations
  with energy-conserving momentum injection}

  \author{ Takashi \textsc{Okamoto}\altaffilmark{1}\altemailmark\orcid{0000-0003-0137-2490}
  \email{takashi.okamoto@sci.hokudai.ac.jp}
  }
  \altaffiltext{1}{Faculty of Science, Hokkaido University, N10 W8 Kitaku, Sapporo 060-0810 Japan}



  \maketitle

\begin{abstract}
Accurate modeling of supernova (SN) feedback in galaxy formation simulations
is complicated by energy conservation violations arising from the vector nature
of momentum injection. We present a mechanical feedback scheme addressing two
key sources: the relative motion between gas elements and the SN-hosting star
particle, and multiple momentum injections into a single gas element within one
timestep. Computing the kinetic energy increment in the rest frame of the gas
element ensures energy conservation while avoiding the momentum inversion that
can occur when this calculation is instead performed in the lab frame.
This correction inherently violates momentum conservation, disturbing the
angular momentum distribution and hindering disk formation when momentum is
coupled on galactic scales. To prevent unphysical large-scale momentum coupling
without an ad hoc maximum coupling radius, we switch to purely thermal feedback
when the cooling radius is resolved by the local inter-element separation.
\add{Our scheme is designed for high- to intermediate-resolution zoom-in simulations
with star particle masses up to $\sim 10^5\,M_\odot$.} Through cosmological
zoom-in simulations of dwarf galaxies ($M_\mathrm{vir} \sim 10^{11}\,M_\odot$)
at two mass resolutions, we demonstrate good convergence in star formation
histories; without the momentum correction, stellar mass in low-resolution runs
falls to as low as 59\% of that in high-resolution counterparts. At the feedback
strength reproducing dwarf galaxy stellar masses, a Milky Way-mass simulation
overproduces stellar mass, suggesting additional processes, such as active
galactic nuclei feedback, are required at this mass scale.
\end{abstract}

\ifsubmission
  \pagewiselinenumbers
\fi
  \section{Introduction}

  \noindent
  The standard model of galaxy formation is based on hierarchical structure formation,
  in which dark matter halos grow through the accretion of matter and mergers
  with other halos. Baryonic gas falls into these halos, cools, and condenses to
  form stars.

  Early numerical simulations of galaxy formation, which primarily modeled the
  effects of gravity and radiative cooling, showed a marked discrepancy with observations
  \citep{nb91}. These simulations predicted that gas would cool too efficiently,
  leading to rapid star formation and resulting in galaxies that were overly
  massive and compact, particularly at early times. This inconsistency, known as
  the ``overcooling problem'' (e.g., \cite{katz92}; \cite{ns00}), pointed to missing
  physical processes. It became clear that additional mechanisms were required
  to regulate star formation by counteracting gravitational collapse and
  radiative cooling.

  Among the various feedback processes considered in galaxy formation theory, supernova
  (SN) feedback is regarded as one of the most important \citep{oka05, aquila}.
  SNe release significant amounts of energy, momentum, and heavy elements into the interstellar
  medium (ISM), producing hot bubbles \citep{mckee1977}, driving turbulence
  \citep{chevalier1985, gent2013}, and powering galactic winds that affect the
  circumgalactic medium (CGM) \citep{martin2005, rupke2005}. As a consequence, SN
  feedback connects stellar evolution with galaxy-scale processes and plays a key
  role in shaping the formation and evolution of galaxies.

  However, accurately modeling supernova feedback in galaxy formation
  simulations remains challenging due to limited numerical resolution. When SN energy
  is deposited as thermal energy without resolving the structure of the SN remnant,
  it is rapidly radiated away, leading to the overcooling problem seen in earlier
  simulations. To address this, various subgrid models have been developed to approximate
  the effects of SN feedback within the resolved scales of the simulation.

  Several authors have proposed models in which the cooling of SN-heated gas is temporarily
  disabled until the gas expands sufficiently \citep{tc01, sti06}. \citet{ds12}
  implemented stochastic thermal feedback, ensuring that the heated gas reaches temperatures
  at which the cooling time is long enough for the energy to have a dynamical impact.
  While such methods enhance the effectiveness of SN feedback, their parameters typically
  require recalibration when the numerical resolution is changed.

  An alternative approach involves stochastically accelerating gas particles that
  receive feedback energy to velocities sufficient for escaping the galaxy.
  These ``wind'' particles are often decoupled from hydrodynamic interactions for a while, allowing
  them to exit star-forming regions with predefined speeds and mass-loading factors
  \citep{sh03}. Although this method can reproduce many observed galaxy properties
  by tuning wind parameters such as velocity and mass loading \citep{ofjt10, okamoto14, illustris},
  it introduces a high degree of phenomenology into the simulations.

  \citet{yepes1997} modelled the unresolved hot phase generated by SN feedback
  by postulating a two-phase medium within individual simulation cells. \citet{sh03}
  implemented a similar two-phase approach in Smoothed Particle Hydrodynamics (SPH)
  simulations. In their formulation, the cold phase dominates the mass, while the
  hot phase dominates the volume, causing each two-phase particle to remain
  gravitationally bound to the galaxy. To drive galactic outflows, \citet{sh03} supplemented
  this framework with decoupled kinetic ``winds'' as an additional feedback
  mechanism.

  \citet{keller2014} proposed an improved two-phase ISM model—often called the ``superbubble’’
  model—within the SPH framework. In this model, particles are explicitly
  partitioned into hot and cold components, and thermal energy from nearby SNe is
  deposited into the hot phase. Crucially, the model incorporates subgrid thermal
  conduction and stochastic evaporation: hot bubbles grow by evaporating
  surrounding cold gas in a physically motivated manner governed by conductive
  heat flux and evaporation rates. This prevents the hot gas from being unrealistically
  diluted and suppresses rapid radiative cooling. As a result, the model remains
  robust across different numerical resolutions and naturally produces galactic outflows.

  Recently, a physically motivated subgrid model known as mechanical feedback
  has been proposed \citep{kimm_cen2014, fire}. This model estimates the
  terminal momentum of supernova remnants (SNRs) based on scaling relations
  derived from high-resolution one-dimensional simulations \citep{chevalier1974, cioffi1988, blondin1998}.
  When the Sedov–Taylor phase is unresolved, the estimated terminal momentum is directly
  imparted to the surrounding gas particles or cells. Mechanical feedback is
  sometimes combined with stochastic thermal feedback to facilitate the
  launching of hot gas outflows \citep{oku22}.

  A major challenge in implementing mechanical feedback lies in the vector nature
  of momentum. Naive implementations often violate the conservation of both
  momentum and energy. \citet{hopkins2018} developed a method that ensures
  conservation of these quantities, even when surrounding gas elements are moving
  relative to the SN-hosting particle. However, this method can still fail to conserve
  energy when a gas element receives feedback from multiple SN events within a single
  timestep, potentially resulting in an overestimate of the injected energy.

  \citet{fire3} addressed this issue by computing the total kinetic energy imparted
  to a gas element by multiple SNe and comparing it with the resulting increase in
  kinetic energy in the lab frame. They then adjusted the magnitude of the
  imparted momentum to ensure that the kinetic energy gain matched the total
  injected energy, thereby correcting energy conservation errors arising from overlapping
  feedback events. However, this adjustment leads to a violation of momentum conservation
  in exchange for correcting energy conservation.

  \citet{chaikin2023} developed a kinetic feedback scheme that conserves not
  only energy, but also linear and angular momentum. Momentum is imparted
  through equal and opposite velocity kicks to pairs of gas particles along
  their line of separation, with kick magnitudes computed based on the particles’
  masses and velocities relative to the star. Statistical isotropy is achieved
  by selecting particle pairs along randomly oriented rays cast from the star.
  To ensure energy conservation, each particle can be kicked only once per timestep,
  which is enforced through a priority system. If either particle in a pair is ineligible
  for a kick, the kick is deferred and the corresponding energy is stored for
  the next timestep. However, such deferments introduce anisotropy in momentum
  injection, as multiple kicks to a gas element within a single timestep tend to
  occur preferentially in low-density directions around SN-hosting particles.
  The postponement also alters the timing of feedback, thereby affecting its dynamical
  consequences. While these limitations may be acceptable in scenarios where
  thermal feedback dominates, as in \citet{chaikin2023}, they may pose challenges
  for mechanical feedback models in which kinetic energy injection plays a central
  role.

In this paper, we adopt an approach similar to that of \citet{fire3},
while addressing certain pathological cases.
\add{Our scheme is designed for high- to intermediate-resolution
zoom-in simulations with star particle masses up to
$\sim 10^5\,M_\odot$.}
We investigate the role of energy constraints in mechanical
feedback by performing a series of galaxy formation simulations.

  The structure of this paper is as follows. In section 2, we describe our new implementation
  of mechanical feedback and provide a brief outline of the simulation setup. In
  section 3, we present the main results. Finally, in section 4, we discuss our
  findings and summarize our conclusions.


  \section{Methodology and simulation setup}
  \label{simulations}
  \noindent
  We implement our feedback model using an earlier version of GIZMO\footnote{Specifically,
  we have modified GIZMO version 2021.} \citep{gizmo}. The code
  computes self-gravity using a solver inherited from GADGET-3
  \citep{aquarius}.
  We employ the meshless finite mass (MFM) method for hydrodynamics, which is implemented in GIZMO.
  In this section, we first provide an overview of the physics implemented in our
  simulations, then describe our new feedback scheme in detail. We also outline the
  simulation setup designed to test our new feedback scheme. Throughout this
  paper, we define the kernel size of a gas element $i$ as
  \begin{equation}
    \frac{4 \pi}{3}h_{i}^{3}\sum_{j}W(\boldsymbol{x}_{j}- \boldsymbol{x}_{i}, h_{i}
    ) = 32, \label{eq:kernel}
  \end{equation}
  where the sum runs over all the gas elements within $h_{i}$ and $W(x, h)$ is the
  cubic-spline kernel function. The kernel size of a star particle is also given
  by equation~(\ref{eq:kernel}).

  \subsection{Gas cooling, star formation, and feedback}

  \change{We calculate radiative cooling and heating with the ultraviolet (UV) background radiation using pre-computed tables, as described in \citet{fire2}. For high-temperature gas ($T \gtrsim 10^4$~K), we use metal-line cooling rates from \citet{wss09}. For low-temperature gas down to 10~K, we incorporate fine-structure and molecular cooling using fits to CLOUDY \citep{cloudy} runs, taken from \citet{fire2}. We also include dust--gas collisional heating and cooling \citep{meijerink_2005}. Self-shielding from the UV background is estimated using the fitting function from \citet{fire2} (see also \cite{rahmati_2013}).}

  We do not impose a pressure floor, as the MFM method avoids artificial
  fragmentation even when the local Jeans length is unresolved, although collapse
  is delayed at insufficient resolution \citep{yamamoto2021}.

  Star formation can occur when the hydrogen number density of a gas element, $n_{\mathrm{H}}$,
  exceeds the threshold density of $n_{\mathrm{th}}= 10^3\,\mathrm{cm}^{-3}$, irrespective
  of the numerical resolution, as in \citet{fire2}. The star formation rate
  density is given by
  \begin{equation}
    \dot{\rho}_{*}= \epsilon_{*}\frac{\rho}{t_{\mathrm{ff}}}, \label{eq:schmidt}
  \end{equation}
  where $\rho$ is the gas density of a gas element, $t_{\mathrm{ff}}= \sqrt{3
  \pi/32 G \rho}$ is the local free-fall time, and $\epsilon_{*}$ is the star
  formation efficiency per free-fall time. Throughout this paper, we adopt
  $\epsilon_{*}= 1$ as in \citet{fire2}, such that star formation is regulated
  via feedback.

  We restrict star formation to gas elements that are self-gravitating \citep{hopkins13b}.
  The local virial parameter for a gas element is estimated as
  \begin{equation}
    \alpha_{\mathrm{vir}}= \frac{15}{4 \pi}\frac{\sigma_{\mathrm{1D}}^{2}+ c_{\mathrm{s}}^{2}}{G
    \rho h^{2}},
  \end{equation}
  where $h$ is the smoothing length of the gas element and $c_{\mathrm{s}}$ is
  the sound speed. The one-dimensional local velocity dispersion,
  $\sigma_{\mathrm{1D}}$, is estimated from the velocities of neighboring gas elements
  within $h$. A gas element is considered star-forming when it has high density
  ($n_{\mathrm{H}}> n_{\mathrm{th}}$) and is gravitationally unstable ($\alpha_{\mathrm{vir}}
  < 1$).

  A star-forming gas element is converted into a star particle during a timestep
  $\Delta t$ with probability
  \begin{equation}
    \mathcal{P}_{*}= 1 - \exp\left(- \frac{\Delta t}{t_{\mathrm{sf}}}\right),
  \end{equation}
  where $t_{\mathrm{sf}}\equiv t_{\mathrm{ff}}/\epsilon_{*}$ is the star
  formation timescale \citep{katz92}. Each star particle represents a simple
  stellar population (SSP) with a Chabrier initial mass function \citep{chabrier03}
  and mass limits of $0.1$--$100\,\msun$.

  After star particles form, they affect the surrounding gas through stellar feedback,
  including stellar winds, core-collapse (CC) and Type Ia supernovae (SNe), mass
  loss from AGB stars, and radiation from massive stars. We adopt the
  metallicity-dependent stellar lifetimes from \citet{pcb98} to compute the timed
  release of mass, metals, and energy from stellar populations. We compute
  stellar wind momenta using tables generated by STARBURST99
  \citep{starburst99}. For CCSNe, we employ the yield tables compiled by \citet{nomoto_2013},
  which assume that stars with masses from 13 to 40~$\msun$ (13 to 300~$\msun$
  for zero-metallicity stars) explode as CCSNe. We lower the minimum mass to the
  canonical value of 8~$\msun$ by extrapolating the tables, which increases the number
  of CCSNe from a star particle by a factor of two compared to the original assumption.
  For Type Ia SNe, we adopt the power-law delay-time distribution from \citet{maoz_2012}
  for simple stellar populations (SSPs) older than $4\times 10^{7}\,\mathrm{yr}$.
  For the chemical yields of Type Ia SNe, we use the metallicity-dependent
  yields of \citet{seitenzahl_2013} (N100 model). For AGB yields, we combine the
  yield tables from \citet{campbell_2008}, \citet{karakas_2010}, \citet{gil-pons_2013},
  and \citet{doherty_2014_a}. We also tabulate the luminosity of the SSP as a function
  of age and metallicity, using the stellar population synthesis code P`EGASE.2 \citep{pegase} to implement radiative feedback.

  We model CCSNe as mechanical feedback, whereas SNe Ia are implemented as
  purely thermal feedback. Mechanical feedback requires treating each supernova event
  individually. In high-resolution simulations, a single star particle may produce
  fewer than one SN Ia over a Hubble time, making it impractical to apply
  mechanical feedback to these events. This thermal treatment is justified because
  SNe Ia typically occur in lower-density environments than CCSNe. We model
  stellar winds and radiation pressure as momentum-driven rather than as
  mechanical feedback.

  \subsection{Mechanical feedback}
  The basic idea of mechanical feedback is to give the terminal radial
  momentum of a gas shell to the surrounding gas when an SN remnant is not
  resolved \citep{kimm_cen2014, fire, hopkins2018}. We employ the terminal
  momentum parameterized by \citet{cioffi1988}:
  \begin{equation}
    \frac{p^{\mathrm{t}}}{\msun \, \mathrm{km}\,\mathrm{s}^{-1}}= 4.8 \times 10^{5}
    \left(\frac{E_{\mathrm{SN}}}{10^{51}\,\mathrm{erg}}\right)^{\frac{13}{14}}\left
    (\frac{n_{\mathrm{H}}}{\mathrm{cm}^{-3}}\right)^{-\frac{1}{7}}f(Z)^{-\frac{3}{14}}
    , \label{eq:terminal}
  \end{equation}
  where $E_{\mathrm{SN}}$ is the energy from an SN event, $n_{\mathrm{H}}$ and $Z$
  are the number density and the metallicity of the ambient gas, respectively, and
  $f(Z)$ is
  \begin{equation}
    f(Z) = \max\left[0.01, \min\left[1, Z/Z_{\odot}\right]\right].
  \end{equation}
  %

  A gas element $j$ is coupled to a star particle $\beta$ if either lies within
  the other's kernel size, $h_{j}$ or $h_{\beta}$. To distribute mass, metals,
  momentum, and energy to neighboring gas elements, we first compute weights
  between the star particle and gas elements, following \citet{hopkins2018}. We do
  not repeat the details here; readers interested in the exact procedure are
  referred to sections 2.2.3 and 2.2.4 of \citet{hopkins2018}.

  We first calculate a scalar weight $\mu_{\beta j}$ with the following property:
  \begin{equation}
    \mu_{\beta j}\simeq \frac{\Delta \Omega_{\beta j}}{4 \pi},
  \end{equation}
  where $\Delta \Omega_{\beta j}$ is the solid angle subtended by the effective
  hydrodynamic interface between the star particle $\beta$ and its neighboring
  gas element $j$. This weight is normalized as
  \begin{equation}
    \tilde{\mu}_{\beta j}= \frac{\mu_{\beta j}}{\sum_{l}\mu_{\beta l}},
  \end{equation}
  where the sum is taken over all neighboring gas elements coupled to this feedback
  event. This ensures
  \begin{equation}
    \sum_{j}\tilde{\mu}_{\beta j}= 1. \label{eq:weight0}
  \end{equation}
  Computing this weight requires only a single pass over the neighboring gas
  elements.

  To ensure momentum conservation, we need vector weights that satisfy the following
  properties:
  \begin{equation}
    \sum_{j}\boldsymbol{w}_{\beta j}= 0, \label{eq:weight}
  \end{equation}
  \begin{equation}
    \sum_{j}|\boldsymbol{w}_{\beta j}| = 1,
  \end{equation}
  and
  \begin{equation}
    |\boldsymbol{w}_{\beta j}| = w_{\beta j}\simeq \tilde{\mu}_{\beta j}\simeq
    \frac{\Delta \Omega_{\beta j}}{4 \pi}.
  \end{equation}
  The absolute value of the vector weight, $w_{\beta j}$, can be used as a
  scalar weight. Computing $\boldsymbol{w}_{\beta j}$ requires two passes over the
  neighboring gas elements.

  \subsubsection{Conservation in a static medium}
  When gas elements are static relative to a star particle, implementing
  feedback that conserves both energy and momentum is straightforward. The ejecta
  mass transferred from a star particle $\beta$ to a gas element $j$ is estimated
  as
  \begin{equation}
    \Delta M_{\beta j}= w_{\beta j}M_{\beta}^{\mathrm{ej}},
  \end{equation}
  where $M_{\beta}^{\mathrm{ej}}$ is the ejecta mass from the star particle $\beta$
  during this timestep\footnote{The subscript $\beta$ indicates quantities
  related to star $\beta$. Therefore, Einstein's summation convention does not
  apply to $\beta$. We use explicit summation symbols throughout this paper to avoid
  confusion.}. The momentum imparted to the gas element $j$ is then
  \begin{equation}
    \Delta \boldsymbol{p}_{\beta j}= \boldsymbol{w}_{\beta j}p_{\beta}, \label{eq:p_simple}
  \end{equation}
  where $p_{\beta}$ is the total radial momentum to be distributed among
  the gas neighbors. The terminal momentum $p_{\beta}^{\mathrm{t}}$ in equation~(\ref{eq:terminal})
  is a natural choice for $p_{\beta}$. However, always using the terminal momentum
  can result in distributing more kinetic energy than is available from the feedback
  event in low-density environments. The momentum added to the
  gas element $j$ is thus usually capped by the maximum possible momentum:
  \begin{equation}
    \Delta p_{\beta j}^{\mathrm{max}}= \sqrt{2 \left(m_{j}+ \Delta M_{\beta j}\right)
    \Delta E_{\beta j}},
  \end{equation}
  where $\Delta E_{\beta j}= w_{\beta j}E_{\beta}^{\mathrm{SN}}$ and $E_{\beta}
  ^{\mathrm{SN}}$ is the total energy available from this event. By using this
  maximum possible momentum, equation~(\ref{eq:p_simple}) becomes
  \begin{equation}
    \Delta \boldsymbol{p}_{\beta j}= \boldsymbol{w}_{\beta j}\min\left[p_{\beta}
    ^{\mathrm{t}}, \frac{\Delta p_{\beta j}^{\mathrm{max}}}{w_{\beta j}}\right]
    . \label{eq:mech1}
  \end{equation}
  Here, $p_{\beta}^{\mathrm{t}}$ is the terminal momentum estimated for this event.
  While this estimation of $\Delta \boldsymbol{p}_{\beta j}$ may slightly
  violate momentum conservation due to neighbor-dependent capping, it ensures that
  the added energy does not exceed $E_{\beta}^{\mathrm{SN}}$.

  \subsubsection{Non-static case: Conservation with relative motion}
  \label{sec:single}

  The above discussion applies only to the static case. In reality, gas elements
  typically have non-zero velocities relative to the star particle, which can easily
  lead to violations of energy conservation (i.e., injecting more energy than is
  available). Determining the appropriate maximum $p_{\beta}$ thus requires
  accounting for the mass and velocity distributions of the gas elements coupled
  to this event. Such a method is presented in Appendix E of \citet{hopkins2018}.
  We take a slightly different approach here.

  We first split the feedback process into two steps: the mass transfer step and
  the energy and momentum transfer step. We complete mass transfer from the star
  particle to its gas neighbors during the weight calculation loops. Since the
  vector weights $\boldsymbol{w}_{\beta j}$ are not available at this stage, we
  use $\tilde{\mu}_{\beta j}$ instead. Therefore, the mass and velocity of the gas
  element $j$ are updated as
  \begin{align}
     & \tilde{\boldsymbol{v}}_{j}= \frac{m_{j}\boldsymbol{v}_{j}+ \sum_{\beta}\Delta m_{\beta j}\boldsymbol{v}_{\beta}}{m_{j}+ \sum_{\beta}\Delta m_{\beta j}}, \\
     & \tilde{m}_{j}= m_{j}+ \sum_{\beta}\Delta m_{\beta j}.
  \end{align}

  To compute the maximum momentum, $p_{\beta}^{\mathrm{max}}$, that is realized
  when $E_{\beta}^{\mathrm{SN}}$ is converted into the kinetic energy of the
  gas, we consider energy conservation in the rest frame of the star particle $\beta$.
  In this frame, the momentum of the gas element $j$ is represented as
  \begin{equation}
    \boldsymbol{p}'_{j}= \tilde{m}_{j}\left(\tilde{\boldsymbol{v}}_{j}- \boldsymbol
    {v}_{\beta}\right),
  \end{equation}
  where the prime indicates quantities measured in the rest frame of the star particle
  $\beta$. The momentum added to this gas element in this frame is the same as
  in the lab frame:
  \begin{equation}
    \Delta \boldsymbol{p}'_{\beta j}= \boldsymbol{w}_{\beta j}p_{\beta}.
  \end{equation}
  The kinetic energy added to the gas element $j$ by this momentum injection is
  \begin{align}
    \Delta{E'}_{\beta j}^{\mathrm{KE}} & = \frac{(\boldsymbol{p}_{j}' + \Delta \boldsymbol{p}'_{\beta j})^{2}}{2 \tilde{m}_{j}}- \frac{{\boldsymbol{p}'_j}^{2}}{2 \tilde{m}_{j}}\nonumber \\
                                        & = \frac{\boldsymbol{w}_{\beta j}^{2}p_{\beta}^{2}+ 2 \boldsymbol{p}'_{j}\cdot\boldsymbol{w}_{\beta j}p_{\beta}}{2 \tilde{m}_{j}}.
  \end{align}
  The total kinetic energy increase from this momentum injection is thus
  \begin{align}
    \Delta{E'}_{\beta}^{\mathrm{KE}} & = \sum_{j}\Delta{E'}_{\beta j}^{\mathrm{KE}}\nonumber                                                                                                       \\
                                      & = p_{\beta}^{2}\sum_{j}\frac{w_{\beta j}^{2}}{2 \tilde{m}_{j}}+ p_{\beta}\sum_{j}\frac{\boldsymbol{p}'_{j}\cdot\boldsymbol{w}_{\beta j}}{\tilde{m}_{j}},
  \end{align}
  where the summation is taken over all gas particles coupled to the star particle $\beta$.
  Given the available energy $E_{\beta}^{\mathrm{SN}}$, the maximum value of
  $p_{\beta}$ is
  \begin{equation}
    p_{\beta}^{\mathrm{max}}= \frac{-b + \sqrt{b^{2}+ a c}}{a},
  \end{equation}
  where
  \begin{align}
    a & = \sum_{j}\frac{w_{\beta j}^{2}}{\tilde{m}_{j}}, \label{eq:a}                                  \\
    b & = \sum_{j}\frac{\boldsymbol{p}'_{j}\cdot\boldsymbol{w}_{\beta j}}{\tilde{m}_{j}}, \label{eq:b}
  \end{align}
  and
  \begin{equation}
    c = 2 E_{\beta}^{\mathrm{SN}}.
  \end{equation}
  Therefore,
  \begin{equation}
    \Delta \boldsymbol{p}'_{\beta j}= \Delta \boldsymbol{p}_{\beta j}= \boldsymbol
    {w}_{\beta j}\min\left[p_{\beta}^{\mathrm{t}}, p_{\beta}^{\mathrm{max}}\right
    ] \label{eq:mom-isolation}
  \end{equation}
  ensures both momentum conservation and that the kinetic energy increment is less
  than or equal to the available energy.
  The computational overhead is an additional loop
  for calculating equations~(\ref{eq:a}) and (\ref{eq:b}).

  \subsubsection{Multiple momentum injections}
  \label{sec:multiple}

  The above discussion is limited to cases where each gas element receives
  momentum from a single star particle. The simplest solution is to process all feedback
  events coupled to the same gas element serially.
  That is, when two or more star
  particles inject momentum into the same gas element, each feedback event is processed
  sequentially using the mass and velocity updated by previous events. However, this
  approach can be computationally too expensive since a single gas element can often
  be coupled to more than ten feedback events in the same timestep.

  We therefore chose an alternative approach in which we adjust the magnitude of the total momentum increment for the gas element
  $j$,
  \begin{equation}
    \Delta \boldsymbol{p}_{j}= \sum_{\beta}\Delta \boldsymbol{p}_{\beta j},
  \end{equation}
  obtained by the method described in section~\ref{sec:single}, such that the kinetic
  energy increment does not exceed the intended value while preserving the direction
  of $\Delta \boldsymbol{p}_{j}$. The momentum added to the gas element $j$ thus
  becomes
  \begin{equation}
    \Delta \boldsymbol{p}_{j}^{*}= \mathcal{R}_{j}\Delta \boldsymbol{p}_{j}.
  \end{equation}
  To obtain the appropriate $\mathcal{R}_{j}$, we compute the kinetic energy increment
  from momentum injections in the rest frame of the gas element $j$ before any momentum
  coupling (moving with $\tilde{\boldsymbol{v}}_{j}$). In this frame, the momentum
  of the gas element before coupling is
  \begin{equation}
    \boldsymbol{p}_{j}'' = 0,
  \end{equation}
  and the momentum added by the feedback event from star particle $\beta$ is
  \begin{equation}
    \Delta \boldsymbol{p}_{\beta j}'' = \Delta \boldsymbol{p}_{\beta j}= \boldsymbol
    {w}_{\beta j}p_{\beta},
  \end{equation}
  where the double prime indicates quantities in the rest frame of the gas element.
  The intended kinetic energy increment is
  \begin{equation}
    \Delta{E''}_{j}^{\mathrm{KE, int}}= \sum_{\beta}\frac{\left(\Delta{\boldsymbol{p}''}_{\beta
    j}\right)^{2}}{2 \tilde{m}_{j}}. \label{eq:intended}
  \end{equation}
  However, the actual increase in the kinetic energy of the gas element $j$ in this
  frame due to multiple momentum injections is
  \begin{equation}
    \Delta{E''}_{j}^{\mathrm{KE}}= \frac{\left(\Delta{\boldsymbol{p}''}_{j}\right)^{2}}{2
    \tilde{m}_{j}}, \label{eq:actual}
  \end{equation}
  where
  \begin{equation}
    \Delta{\boldsymbol{p}''}_{j}\equiv \sum_{\beta}\Delta \boldsymbol{p}''_{\beta
    j}.
  \end{equation}
  The ratio of equation~(\ref{eq:intended}) to equation~(\ref{eq:actual}) gives $\mathcal{R}
  _{j}$ as
  \begin{equation}
    \mathcal{R}_{j}= \sqrt{\frac{\Delta{E''}_{j}^{\mathrm{KE, int}}}{\Delta{E''}_{j}^{\mathrm{KE}}}}
    .
  \end{equation}
  We update the velocity of the gas element $j$ as
  \begin{equation}
    \boldsymbol{v}_{j}\rightarrow
    \begin{cases}
      \tilde{\boldsymbol{v}}_{j}+ \mathcal{R}_{j}\dfrac{\Delta \boldsymbol{p}_j}{\tilde{m}_j} & \text{for }\mathcal{R}_{j}< 1, \\
      \tilde{\boldsymbol{v}}_{j}+ \dfrac{\Delta \boldsymbol{p}_j}{\tilde{m}_j}                & \text{otherwise},
    \end{cases}
  \end{equation}
  where
  \begin{equation}
    \Delta \boldsymbol{p}_{j}= \sum_{\beta}\Delta \boldsymbol{p}_{\beta j}.
  \end{equation}
  When $\mathcal{R}_{j}> 1$, we add the residual energy, $\Delta{E''}_{j}^{\mathrm{KE,
  int}}- \Delta{E''}_{j}^{\mathrm{KE}}$, to the gas element $j$ in the form of
  thermal energy. \citet{fire3} present a similar approach for the multiple-injection
  case. They, however, calculate the energy change in the lab frame, in which
  case $\mathcal{R}_{j}$ can be negative. Our approach avoids such a pathological
  situation.

  \subsubsection{Avoiding large-scale momentum coupling}

  The method described above can violate momentum conservation.
  A notable numerical artifact we identified is that simulations employing this feedback scheme tend to fail
to form galactic disks.
This appears to occur when the coupling scale is allowed to become
comparable to or larger than the galactic scale.
Furthermore, when the coupling
radius becomes excessively large, momentum is instantaneously transferred to
distant gas elements, which is itself unphysical. A common practice to mitigate
this issue is to introduce a maximum coupling radius of a few kpc \citep{smith2018, fire3}.
When this radius is sufficiently smaller than the galactic scale,
the directions of the spurious residual momentum arising from
the conservation violation would be effectively randomized on
galactic scales, so that its net effect on the large-scale
dynamics of the galaxy would be small.

  However, introducing a maximum coupling radius creates several undesirable behaviors.
  First, as the number of coupled gas elements decreases, the feedback becomes
  increasingly anisotropic, departing from the spherically symmetric assumption.
  Second, when only a single gas element lies within the maximum radius, there
  is no way to conserve momentum for that feedback event, as momentum conservation
  requires at least two gas elements to balance the momentum injection. Third, it
  is possible that no gas elements exist within the maximum radius, in which
  case the feedback energy and momentum are either lost or must be handled
  through ad hoc procedures. Finally, the maximum radius itself becomes a
  critical parameter that significantly affects simulation outcomes. In our simulations,
  we found that decreasing the maximum radius resulted in stronger feedback
  effects, suggesting that this parameter can substantially alter the physical behavior
  of the system in ways that are difficult to predict a priori.

  We therefore take an alternative approach. During the weight calculation loop,
  each star particle computes the cooling radius for its nearest gas element,
  which is given by
  \begin{equation}
    r_{\mathrm{cool}}= 28.4~\mathrm{pc}\left(\frac{n_{\mathrm{H}}}{\mathrm{cm}^{-3}}
    \right)^{-3/7}\left(\frac{E_{\mathrm{tot}}}{10^{51}~\mathrm{erg}}\right)^{2/7}
    f(Z)^{-1/7}, \label{eq:rcool}
  \end{equation}
  where $E_{\mathrm{tot}}$ is the total energy including both the internal
  energy of the gas element and the SN energy from the star particle, $E_{\mathrm{SN}}$.
  This equation has the same form as that in \citet{hopkins2018}, but they use
  $E_{\mathrm{SN}}$ instead of $E_{\mathrm{tot}}$. Such a treatment is valid when
  an explosion occurs in a cold medium where the ambient pressure is negligible.
  In practice, however, an explosion can occur in an already hot medium. Since equation~(\ref{eq:terminal}),
  which is used to derive the cooling radius, is only applicable to an explosion
  in a pressureless medium, we include the internal energy of the gas element to
  account for explosions in hot media as well.

  When the cooling radius is resolved, i.e.,
  \begin{equation}
    2 \bar{h}< r_{\mathrm{cool}},
  \end{equation}
  where $\bar{h}$ is the mean separation of gas elements around this gas element,
  we inject all the SN energy from the star particle into this nearest gas element
  in the form of thermal energy. Both momentum and energy conservation are then ensured by hydrodynamics, with information propagating at an appropriate velocity rather than being instantaneously coupled to distant elements.  Since $r_{\mathrm{cool}}\propto n_{\mathrm{H}}^{-3/7}$
  while $\bar{h}\propto n_{\mathrm{H}}^{-1/3}$, this criterion naturally
  transitions to thermal feedback in low-density regions without imposing an ad
  hoc maximum coupling radius. It also switches to pure thermal feedback when an
  explosion occurs in an already hot medium.

  \subsubsection{Other feedback mechanisms}

  In addition to SN feedback, our simulations include stellar winds,
  photoionization heating, and radiation pressure as sources of stellar feedback.
  For stellar winds, we tabulate the wind momentum as functions of the age and metallicity
  of the stellar population using STARBURST99 \citep{starburst99}.
  This momentum is distributed to the coupled gas elements using the vector weight
  $\boldsymbol{w}_{\beta j}$.

  Young stellar populations also emit ionizing radiation. We compute the number
  of ionizing photons emitted by each star particle during a timestep using look-up
  tables generated by P\'{E}GASE.2 \citep{pegase}. For each gas
  element within the coupling radius of the star particle, we stochastically determine
  whether it is ionized based on its density and the available ionizing photons,
  as in \citet{marinacci2019}, without explicitly solving the radiation transfer
  equation. Ionized gas elements are heated to $10^{4}$~K and forbidden to cool below
  $10^{4}$~K until the end of the star particle's timestep or the occurrence of its
  first supernova, whichever comes first.

  Radiation from young star particles also impacts the surrounding gas through
  radiation pressure \citep{hopkins11, agertz13, okamoto14, ishiki_2017}. The
  total momentum that a star particle $\beta$ injects into its coupled gas
  elements is
  \begin{equation}
    p_{\beta}^{\mathrm{rad}}= \frac{L_{\beta}^{\mathrm{UV}}}{c}\Delta t_{\beta}
    , \label{eq:p_rad}
  \end{equation}
  where $L_{\beta}^{\mathrm{UV}}$ and $\Delta t_{\beta}$ are the UV luminosity
  and the timestep of the star particle, respectively. In equation~(\ref{eq:p_rad}),
  we do not include the effect of multiple scattering by dust grains, which is often
  included in simulations (e.g., \cite{horie_2024}), since the increase in
  radiation pressure from this effect compensates for the reduction in
  photoionization heating by dust \citep{ishiki_2017}.

  \subsubsection{Timestep constraints for feedback}

  Since the terminal momentum given by equation~(\ref{eq:terminal}) is valid only for
  a single SN event, we restrict the timestep of a young star particle so that at most
  one SN event occurs in a single timestep of that particle. When the number of SNe
  expected from a star particle since its last feedback event (or formation, if no
  feedback has occurred yet) is less than one, we postpone the feedback until the
  expected number reaches unity or the age of the star particle exceeds the lifetime
  of an $8~\msun$ star, whichever comes first.

  The timesteps of the coupled gas elements should be as short as that of the SN
  host star particle; otherwise, these gas elements do not recognize that they
  receive momentum injection. Furthermore, we restrict the timestep of each gas
  element so that it does not exceed twice the timestep of any of its
  neighboring gas elements with which it has hydrodynamic interactions \citep{sm09, durier_2012}\footnote{Such
  a timestep limiter is already implemented in GIZMO, but it is based
  on the signal velocity. Since the timestep limited by stellar evolution does
  not affect the signal velocity, we modify the code so that the timestep of a gas
  element is directly limited by the timesteps of its neighboring gas elements and
  young star particles.}.

  Due to these restrictions, low-resolution simulations can be computationally very
  expensive since larger star particles host more SN events per unit time.
  Consequently, our method is impractical for low-resolution simulations where the
  star particle mass exceeds $10^{5}\,\msun$. We have confirmed that relaxing
  these timestep criteria generally results in stronger feedback effects at
  lower resolutions and worsens the numerical convergence.

  \subsection{Simulations}

  In this paper, we perform cosmological simulations assuming the $\Lambda$-dominated
  cold dark matter ($\Lambda$CDM) model. The cosmological parameters are: $\Omega
  _{0}= 0.308$, $\Omega_{\Lambda}= 0.692$, $\Omega_{\mathrm{b}}= 0.0484$,
  $H_{0}= 67.81\,\mathrm{km}\,\mathrm{s}^{-1}\,\mathrm{Mpc}^{-1}$, $\sigma_{8}= 0
  .815$, and $n_{\mathrm{s}}= 0.968$, consistent with \citet{planck15}.

  To investigate numerical convergence, we simulate two dwarf galaxies with halo
  masses of $M_{\mathrm{vir}}\simeq 10^{11}\,\msun$ at $z = 0$, each at two
  different resolutions. To generate initial conditions for these simulations,
  we use MUSIC \citep{music}. We first run a dark-matter-only
  simulation in a periodic cubic volume with a comoving side length of 25~Mpc. We
  then randomly select two isolated dark matter halos at $z = 0$ with virial masses
  of $\simeq 10^{11}\,\msun$. For each halo, we generate a new initial condition
  by adding shorter-wavelength perturbations to the Lagrangian region from which
  the selected halo forms. The two dwarf galaxies are called DW1 and DW2, respectively,
  and high- and low-resolution simulations are distinguished by HR and LR,
  respectively. In the low-resolution simulations, the dark matter particle mass
  and the initial gas element mass in the zoom region are
  $m_{\mathrm{DM}}\simeq 4.8 \times 10^{5}\,\msun$ and
  $m_{\mathrm{gas}}\simeq 9.0 \times 10^{4}\,\msun$, respectively. In the high-resolution
  simulations, these masses are $m_{\mathrm{DM}}\simeq 6.0 \times 10^{4}\,\msun$
  and $m_{\mathrm{gas}}\simeq 1.1 \times 10^{4}\,\msun$.

  For the dark matter particles in the zoom regions, we employ Plummer-equivalent
  softening lengths of $\epsilon_{\mathrm{DM}}= 3.48$~ckpc in comoving
  coordinates, switching to a fixed physical softening of $\epsilon_{\mathrm{DM,
  max}}= 0.348$~kpc at $z = 9$ for LR simulations. For HR simulations, these values
  are $\epsilon_{\mathrm{DM}}= 1.23$~ckpc and $\epsilon_{\mathrm{DM, max}}= 0.123$~kpc.
  For gas elements, we employ adaptive gravitational softening without imposing a
  minimum softening length.

  We also simulate a Milky Way-mass galaxy to investigate the galaxy-mass dependence
  of our feedback scheme. For this simulation, we use a simulation box with a comoving
  side length of 50~Mpc. The mass resolution is the same as in the LR
  simulations, and the gravitational softening lengths are $\epsilon_{\mathrm{DM}}
  = 1.99$~ckpc and $\epsilon_{\mathrm{DM, max}}= 0.199$~kpc. These softening
  lengths are determined following \citet{fire2}. This simulation is referred to
  as MW.

  To investigate how our feedback scheme works for different feedback strength, we
  change the energy per supernova as $E_{\mathrm{SN}}= \eta_{\mathrm{SN}}\times 1
  0^{51}$~erg. In this paper, we show the results with $\eta_{\mathrm{SN}}= 2, 3,$
  and $4$.

  \section{Results}

  In this section, we present the results of our new feedback scheme, organized
  into three primary investigations. First, we investigate the dependence of our
  results on numerical resolution by simulating a suite of dwarf galaxies. Second,
  we demonstrate the importance of imposing a stringent limit on kinetic energy
  increment by disabling the momentum correction described in section
  \ref{sec:multiple}. Finally, we show a simulation of a Milky Way-mass galaxy
  to investigate the properties of our feedback scheme on a different mass scale.

  \subsection{Global properties of dwarf galaxies}
  \label{sec:dwarf}

  \begin{figure}[h!]
    {\centering \includegraphics[width=8.5cm]{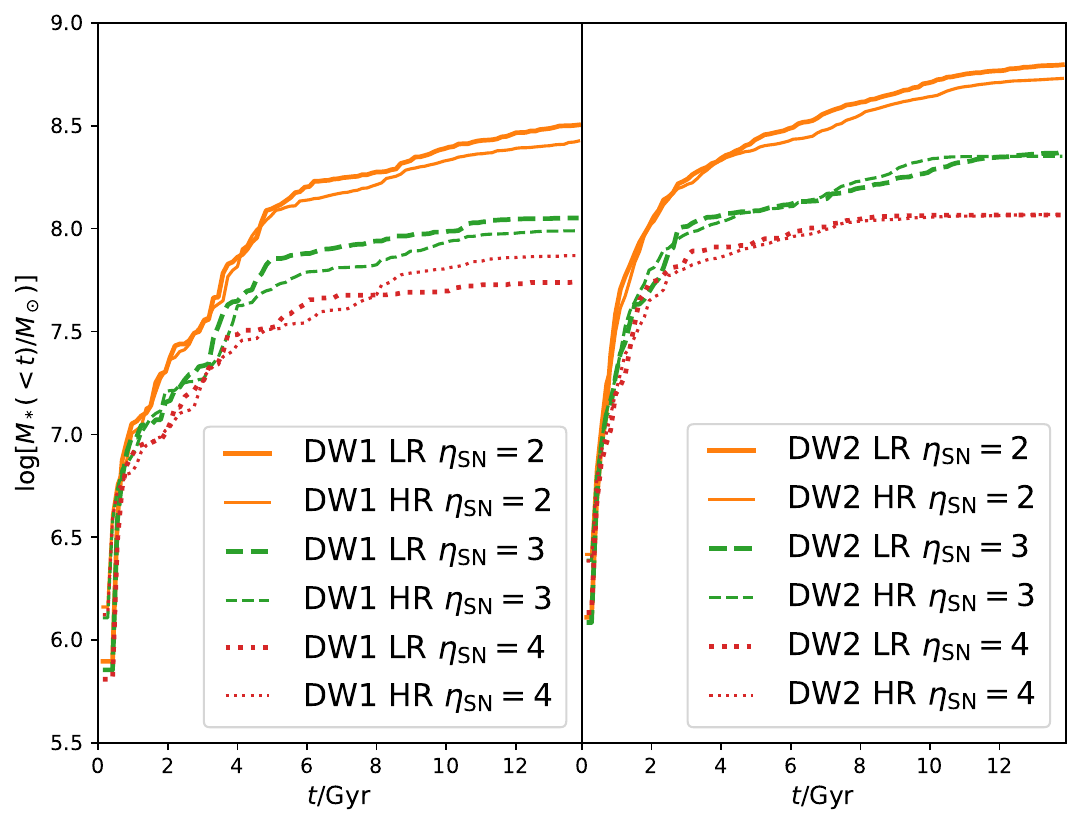} }
    \caption{ Star formation history (cumulative stellar mass growth) of stars within
    the virial radii of DW1 (left panel) and DW2 (right panel) halos at $z=0$.
    The vertical axis shows the cumulative stellar mass formed up to cosmic age
    $t$ for stars that reside within the virial radius at $z=0$. Thick and thin lines
    indicate high-resolution (HR) and low-resolution (LR) simulations,
    respectively. Orange solid, green dashed, and red dotted lines correspond to
    the supernova energy efficiency factor,
    $\eta_{\text{SN}}=2, 3, \text{and }4$, respectively.
    \alttxt{Line plots of cumulative stellar mass growth versus cosmic age
for two dwarf galaxies at two resolutions and three feedback strengths.}
    \label{fig:sfh} }
  \end{figure}
  We first compare the \add{cumulative} star formation histories of DW1 and DW2 by varying resolution
  and feedback strength. As shown in figure~\ref{fig:sfh}, the results are
  insensitive to the adopted resolution. The largest difference in stellar mass between
  different resolution simulations occurs for DW1 with $\eta_{\mathrm{SN}}= 4$.
  Even in this case, the difference at $z = 0$ is only a factor of 1.34. As
  expected, the stellar mass decreases with increasing feedback strength.

  %
  \begin{figure}[h!]
    {\centering \includegraphics[width=8.5cm]{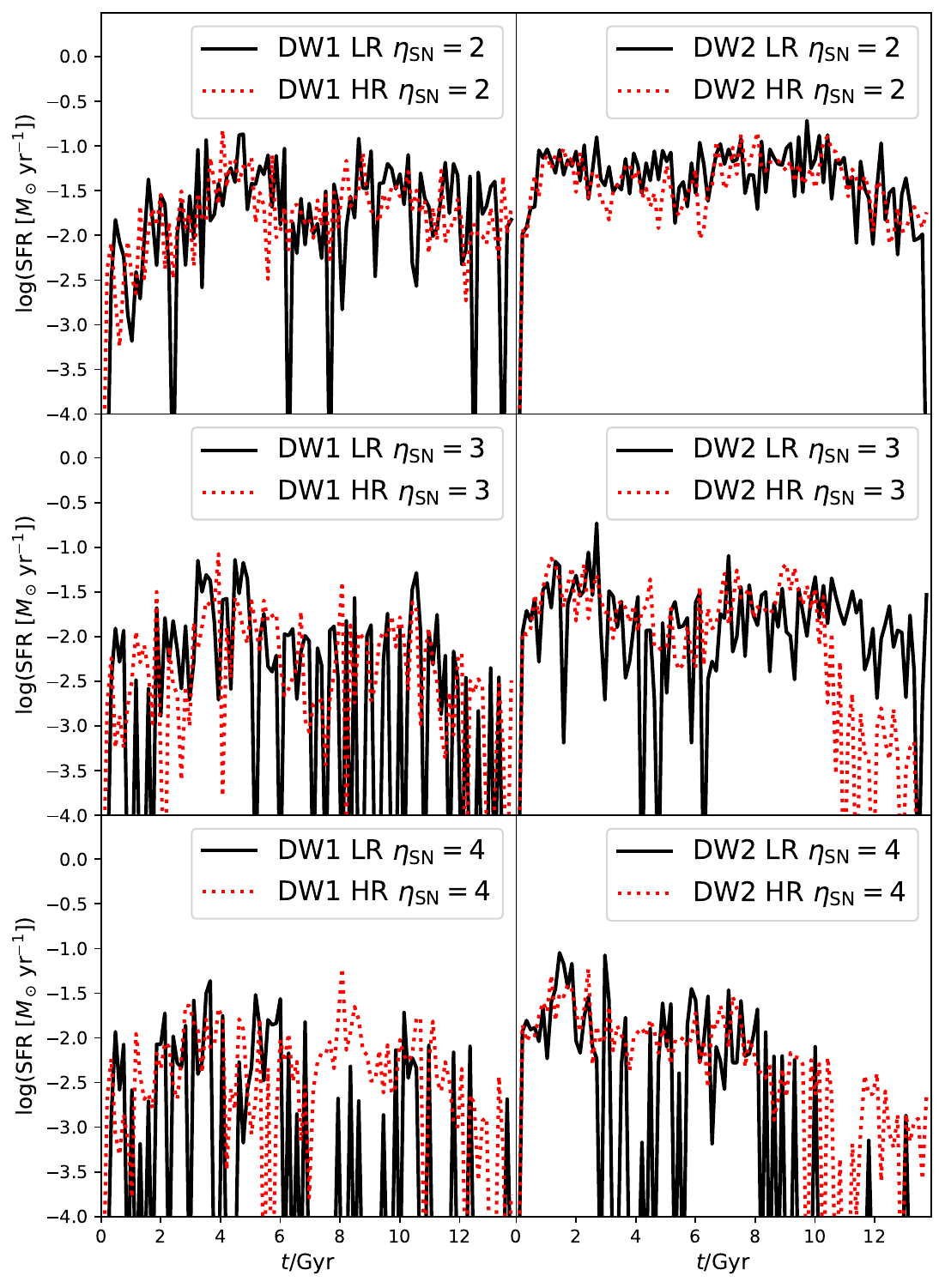} }
    \caption{Star formation rate history for DW1 (left panels) and DW2 (right panels)
    galaxies. The vertical axis shows the star formation rate (SFR) within the galactic
    radius (see the text). Each row corresponds to a different supernova
    feedback efficiency factor ($\eta_{\text{SN}}=2, 3, \text{and }4$ from top to
    bottom). Solid black lines indicate low-resolution (LR) simulations, and red
    dotted lines indicate high-resolution (HR) simulations.
    \alttxt{Line plots of star formation rate history versus cosmic age for
two dwarf galaxies at two resolutions and three feedback strengths.}
    \label{fig:sfr} }
  \end{figure}
  figure~\ref{fig:sfr} presents the star formation rate (SFR) histories within the
  galactic radius for DW1 (left panels) and DW2 (right panels) halos, examining
  the impact of resolution and feedback strength. Here, the galactic radius is
  defined as twice the stellar half-mass radius at $z = 0$ to exclude extended
  stellar halos from our analysis. In contrast to the smooth cumulative stellar mass
  growth shown in figure~\ref{fig:sfh}, the SFR histories reveal highly bursty star
  formation behavior characteristic of dwarf galaxies, with fluctuations
  spanning up to several orders of magnitude.

  The impact of resolution on SFR variability depends on feedback strength. For weak
  feedback ($\eta_{\mathrm{SN}}= 2$, top panels), high-resolution (red dotted
  lines) and low-resolution (black solid lines) simulations exhibit similar levels
  of burstiness. However, at stronger feedback efficiencies ($\eta_{\mathrm{SN}}=
  3$
  and 4, middle and bottom panels), LR simulations become more susceptible to quenching,
  showing more frequent and deeper suppression of star formation compared to
  their HR counterparts. This is particularly evident for $\eta_{\mathrm{SN}}= 4$
  (bottom panels), where LR simulations display extended quiescent periods with
  SFR dropping by three to four orders of magnitude. This enhanced quenching efficiency
  in LR runs likely arises from the coarser spatial discretization, where
  feedback energy deposited into larger volumes more effectively evacuates gas
  from star-forming regions, leading to more complete but temporary cessation of
  star formation. This result is consistent with earlier findings by \citet{fire2}.

  An interesting exception to this pattern occurs in DW2 with $\eta_{\mathrm{SN}}
  = 3$ at late times ($t \gtrsim 10$~Gyr). Here, the HR simulation shows strong quenching
  while the LR counterpart continues forming stars. We trace this difference to
  a starburst at $t \simeq 9$~Gyr in the HR run triggered by a minor merger
  event, which subsequently quenches star formation through vigorous feedback-driven
  outflows.
  \change{This merger is not captured in the LR simulation, likely
due to under-resolved dynamical friction or artificial
tidal disruption at lower resolution.}

  Despite these resolution-dependent variations in the SFR histories, the integrated
  stellar masses (figure~\ref{fig:sfh}) remain remarkably consistent between HR
  and LR runs. This convergence suggests that while the timing and intensity of individual
  star formation episodes depend sensitively on resolution, the overall stellar
  mass assembly is primarily regulated by the balance between gas accretion and
  feedback-driven outflows, which our feedback prescription captures in a manner
  that is largely independent of resolution.

  \begin{figure}[h!]
    {\centering \includegraphics[width=8.5cm]{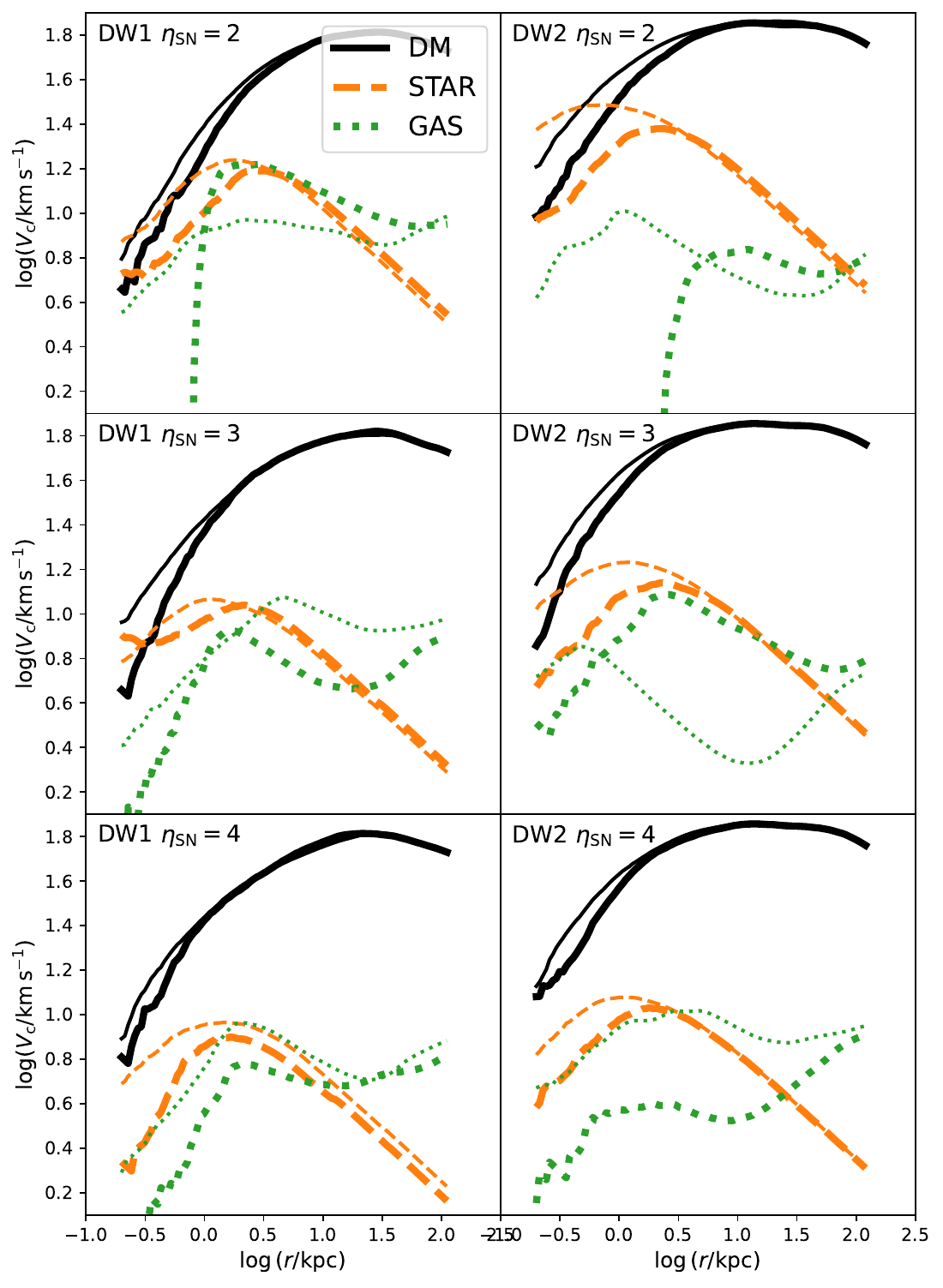} }
    \caption{Circular velocity curves for individual components of DW1 (left column)
    and DW2 (right column) at $z=0$. The vertical axis shows the logarithm of the
    circular velocity contribution,
    $\log_{10}(V_{\mathrm{c}, i}/\text{km s}^{-1})$, where $V_{\mathrm{c}, i}(r)
    = \sqrt{G M_{i}(<r)/r}$, where $i$ can be either dark matter (DM), stars, or
    gas. The contribution of the DM, stars, and gas components are shown by black
    solid, dashed orange, and dotted green lines, respectively. The LR simulations
    are shown by the thick lines and HR counterparts by the thin lines. The
    panels from top to bottom correspond to the supernova energy efficiency
    factors $\eta_{\text{SN}}=2, 3, \text{and }4$.
    \alttxt{Circular velocity curves of dark matter, stellar, and gas
components versus galactic radius for two dwarf galaxies at two resolutions
and three feedback strengths.}
    \label{fig:vc} }
  \end{figure}

  Stellar mass is, of course, one of the most important properties of galaxies. However,
  achieving identical stellar masses (or star formation histories) is
  insufficient evidence for simulation convergence. \citet{crain_2023} showed
  that simulations with feedback recipes that all produce similar stellar mass
  functions exhibit significant differences in the circumgalactic medium (CGM). We
  therefore compare the matter distributions within the virial radius between the
  two different resolution simulations.

  We now compare the internal mass distribution, characterized by the circular
  velocity curves, to assess the convergence of the DW1 and DW2 simulations across
  different resolutions and feedback strengths in figure~\ref{fig:vc}. Unlike the star formation
  history (SFH), which showed relatively robust convergence in the final stellar
  mass, the circular velocity curves reveal significant resolution-dependent
  differences in the inner density profiles of the individual components.

  The convergence quality varies markedly between the baryonic and dark matter (DM)
  components. The DM component (black solid lines) shows the best overall convergence,
  with the high-resolution (HR, thin lines) and low-resolution (LR, thick lines)
  curves closely tracking one another across all feedback strengths for both DW1
  and DW2. This indicates that the gravitational backbone of the halo is robustly
  resolved even with the LR setup. In contrast, the gas (dotted green lines) and
  star (dashed orange lines) components show substantial non-convergence,
  particularly in the inner regions ($\log(r/\text{kpc}) < 0.5$). The gas profiles
  are highly sensitive to resolution, with the LR runs consistently showing
  lower central circular velocities compared to the HR runs, indicating a less centrally
  concentrated gas distribution. For the DW1 halo, increasing feedback strength to
  $\eta_{\text{SN}}=4$ exacerbates the difference, leading to a poorer match
  between the LR and HR star and gas velocity profiles.

  The magnitude of the non-convergence is clearly modulated by the strength of
  the supernova feedback, though the specific response is highly halo-dependent.
  For the DW1 halo, the stellar component profiles (STAR) are reasonably well-matched
  at the mildest feedback ($\eta_{\text{SN}}=2$). However, the DW2 halo exhibits
  substantial non-convergence in the stellar profile even at the mildest feedback
  level ($\eta_{\text{SN}}=2$). For both halos, the low-resolution runs consistently
  predict lower central stellar masses (indicated by lower
  $V_{\mathrm{c}, \text{star}}$ peaks) compared to their HR counterparts. The
  stellar distribution converges at larger radii
  $\log(r/\text{kpc}) \gtrsim 0.5$, suggesting the stellar distribution's dependence
  on numerical resolution is minimal where the dark matter distribution is converged.
  On the other hand, gas distribution is markedly different between resolutions in
  both halos and at all feedback strengths. Gas in the lower resolution
  simulations is more extended than in the higher resolution counterparts. The circular
  velocity curves reach almost the same values at virial radii (the outermost
  radii). This suggests that our feedback is spuriously stronger in the lower
  resolution simulations at galactic scales, while it is resolution-independent at
  the halo scale. Our results indicate that convergence in structure is much
  harder to achieve than convergence in integrated quantities like stellar mass.

  \begin{table}[h]
    \caption{Baryon fractions normalized by the cosmic baryon fraction, $\Omega_{\mathrm{b}}
    /\Omega_{0}$, within the virial radius of the DW1 and DW2 halos at $z=0$. \label{tab:baryon_fractions}
    }
    \centering
    \begin{tabular}{lcc}
      \hline
                                    & LR     & HR     \\
      \hline
      DW1 ($\eta_{\mathrm{SN}}= 2$) & 0.198  & 0.202  \\ 
      DW1 ($\eta_{\mathrm{SN}}= 3$) & 0.147  & 0.205  \\
      DW1 ($\eta_{\mathrm{SN}}= 4$) & 0.0997 & 0.135  \\
      DW2 ($\eta_{\mathrm{SN}}= 2$) & 0.116  & 0.116  \\ 
      DW2 ($\eta_{\mathrm{SN}}= 3$) & 0.0878 & 0.0703 \\
      DW2 ($\eta_{\mathrm{SN}}= 4$) & 0.131  & 0.156  \\
      \hline
    \end{tabular}
  \end{table}

  The baryon fraction within the virial radius is a good measure of feedback strength.
  We summarize the normalized baryon fraction,
  \[
    \frac{M_{\mathrm{star}}+ M_{\mathrm{gas}}}{M_{\mathrm{vir}}}\, \left(\frac{\Omega_{\mathrm{b}}}{\Omega_{0}}
    \right)^{-1},
  \]
  of each simulation at $z = 0$ in table~\ref{tab:baryon_fractions}. We find
  that feedback is very efficient to eject gas from the halos in all simulations;
  all halos lost almost 80\% or more baryons. Overall, the baryon fraction is
  smaller in the LR simulations than in the HR counterparts, suggesting that our
  feedback scheme is slightly stronger in lower resolution even at halo scale.

  We also find that the baryon fraction is not monotonically decreasing with increasing
  feedback strength, $\eta_{\mathrm{SN}}$. Specifically, for the DW2 halo (in
  both LR and HR runs) and the DW1 HR run, the baryon fraction decreases from
  $\eta_{\mathrm{SN}}=2$ to $\eta_{\mathrm{SN}}=3$, but then unexpectedly
  increases when $\eta_{\mathrm{SN}}$ is raised further to 4. The DW2 halo (HR
  run) provides a clear example of this non-monotonic behavior, where the fraction
  drops from 0.116 ($\eta_{\mathrm{SN}}=2$) to 0.0703 ($\eta_{\mathrm{SN}}=3$) but
  rises to 0.156 ($\eta_{\mathrm{SN}}=4$). The non-monotonic behavior is
  explained by the time variability of the gas mass in the virial radius. The gas
  mass decreases when intense feedback occurs after a starburst and increases when
  gas re-accretes onto the halo. The low baryon fractions demonstrate that our
  feedback scheme is capable of ejecting a large amount of gas from the halo
  while successfully maintaining a total stellar mass that is largely
  insensitive to the adopted numerical resolution.
  \subsection{The importance of momentum correction for multiple momentum
  injections}
  \label{sec:non-conserving}

  In this subsection, we run the same dwarf galaxy simulations presented in section
  \ref{sec:dwarf} but without the momentum correction for multiple momentum injections
  described in section \ref{sec:multiple}. While we focus only on simulations
  with $\eta_{\mathrm{SN}}= 3$, the results are qualitatively the same for other
  feedback strengths. Note that even without the momentum correction for
  multiple injections, the overall feedback scheme still conserves both momentum
  and energy in an isolated explosion. The feedback becomes purely thermal when
  the cooling radius is resolved—specifically, when the density of the pre-coupled
  gas is sufficiently low and/or the temperature of the pre-coupled medium is sufficiently
  high.

  \begin{figure}[h!]
    {\centering \includegraphics[width=8.5cm]{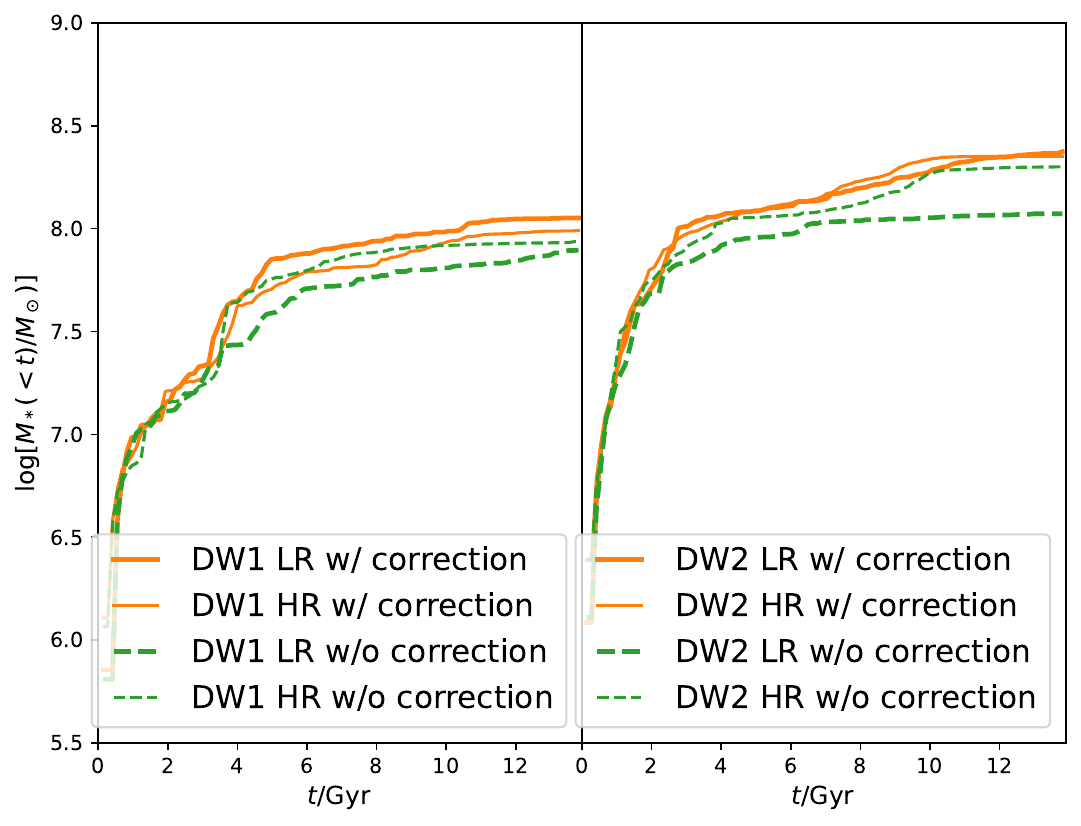} }
    \caption{The same as figure~\ref{fig:sfh}, but comparing the fiducial feedback
    model (orange solid lines) with the one without momentum
    correction for multiple momentum injections (green dashed lines). All simulations
    employ $\eta_{\text{SN}}= 3$. Thick lines correspond to the LR simulations
    and thin lines to the HR simulations.
    \alttxt{Line plots of cumulative stellar mass growth comparing simulations
with and without momentum correction for multiple supernova injections.}
    \label{fig:sfh_noncons} }
  \end{figure}

  In figure~\ref{fig:sfh_noncons}, we compare the star formation histories
  obtained by our fiducial feedback scheme and the model without momentum
  correction. We find that feedback is significantly stronger in the simulations
  without the momentum correction than in our fiducial runs. This suggests that
  without the correction, the kinetic energy increment added to gas elements is often
  larger than the energy physically available for the feedback event.

  In the DW2 halo specifically, the simulation without momentum correction
  exhibits strong resolution dependence: the LR run forms only 59\% of the
  stellar mass produced in its HR counterpart. This level of non-convergence is notably
  worse than the most extreme case observed with our fiducial scheme. Because the
  issue of multiple momentum injections is most pronounced following intense
  starbursts, the impact is highly sensitive to the star formation history of the
  individual halo. This explains why the importance of the momentum correction
  varies between DW1 and DW2.

  The HR simulations without momentum correction show relatively
good agreement with our fiducial feedback results. This is
partly because the transition from mechanical to pure-thermal
feedback occurs earlier at higher resolution. When multiple
momentum injections lead to spuriously strong feedback in
certain events, the resulting intense energy injection
significantly heats and rarefies the local interstellar medium.
This creates conditions---high temperature and low density---where
the cooling radius of subsequent supernova events is more easily
resolved. Consequently, following feedback events are more likely
to transition into the pure-thermal regime, which does not suffer
from the same spurious over-strengthening as the uncorrected
momentum injections. This self-regulating behavior occurs earlier
in the HR simulations. \add{Furthermore, early feedback processes
such as photoionization pre-process the ISM, reducing the gas
density around star-forming regions. This also lowers the
frequency of multiple momentum injections by suppressing
clustered star formation as suggested by \citet{benitez-llambay2026}, which
showed that the inclusion of early feedback processes improves
numerical convergence. Therefore, the inclusion of early feedback
before SNe may also contribute to the convergence of the HR
simulations without momentum correction to the fiducial HR runs.}

  \subsection{A simulation of Milky Way-mass galaxy}
  \label{sec:mw}

  \begin{figure}[h!]
    {\centering \includegraphics[width=8.5cm]{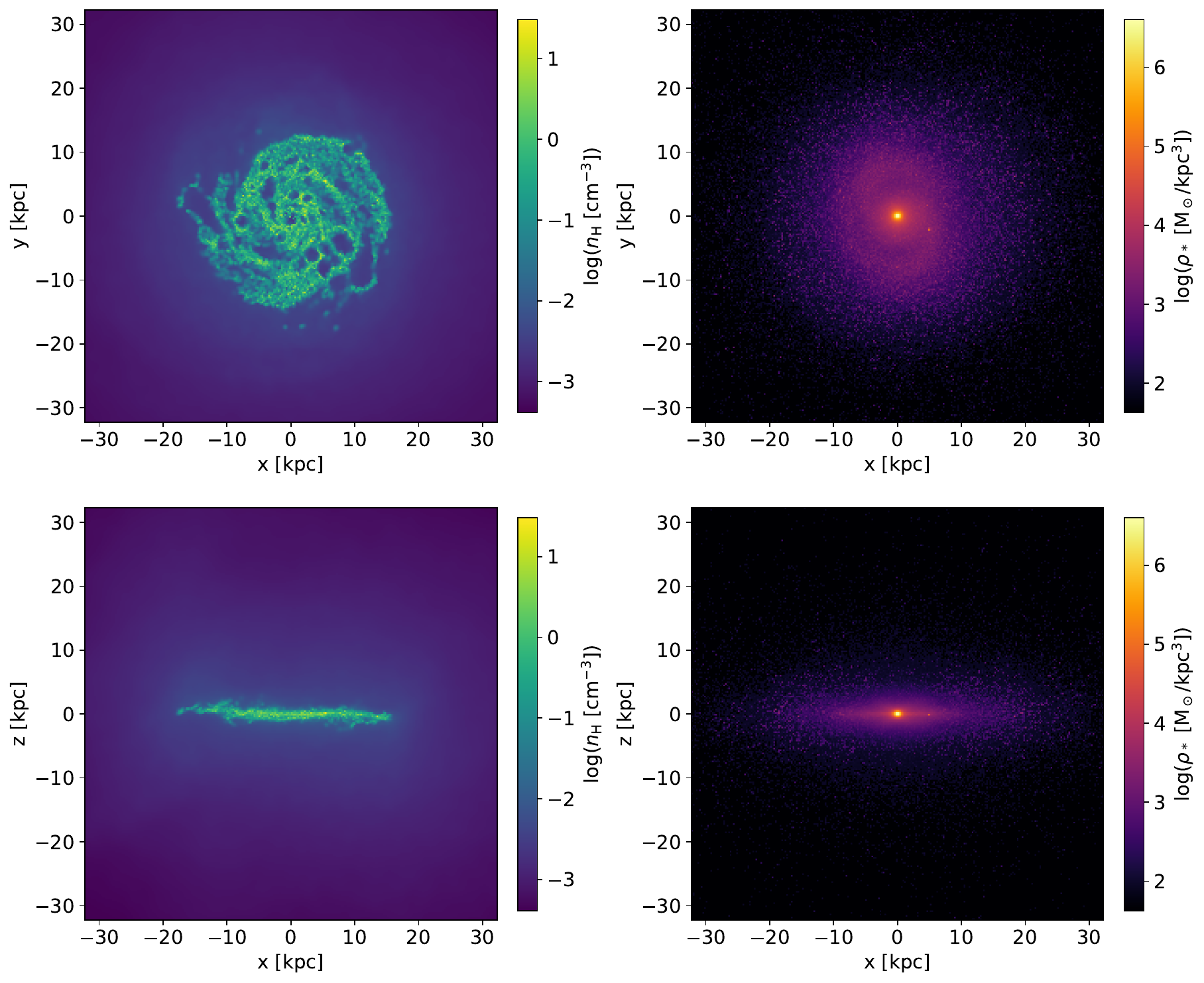} }
    \caption{Density-weighted projected density maps of the Milky Way-mass
    galaxy at $z = 0$. The left and right columns show gas and stellar densities,
    respectively, while the upper and lower rows show face-on and edge-on projections.
    The side length of the region shown is $0.2 R_{\mathrm{vir}}$.
    \alttxt{Projected density maps of a simulated Milky Way-mass galaxy
showing gas and stellar distributions in face-on and edge-on projections.}
    \label{fig:map}
    }
  \end{figure}

In this subsection, we present the results of the MW
simulation. The mass resolution is identical to that of the LR dwarf
galaxy simulations, while the virial mass at $z = 0$ is approximately
$1.8 \times 10^{12}\,M_{\odot}$. For this simulation, we adopt a
feedback strength of $\eta_{\mathrm{SN}} = 3$.

In figure~\ref{fig:map}, we show the gas and stellar distributions at $z = 0$.
The projections demonstrate that the simulation successfully produces a
disk-dominated galaxy with a clearly defined spiral structure. Notably, the
face-on gas distribution reveals numerous kpc-scale holes and cavities
sculpted by intense SN feedback. However, the baryon fraction within the
virial radius at $z = 0$ remains very close to the cosmic baryon fraction.
This indicates that while our feedback is effective at displacing gas from
the ISM, it does not provide sufficient energy to
eject the gas beyond the virial radius at this mass scale. Furthermore, the
edge-on projections confirm the formation of thin stellar and
gas disks, suggesting that our feedback scheme maintains a stable, cold
disk environment while simultaneously driving gaseous outflows within the halo.

  \begin{figure}[h!]
    {\centering \includegraphics[width=8.5cm]{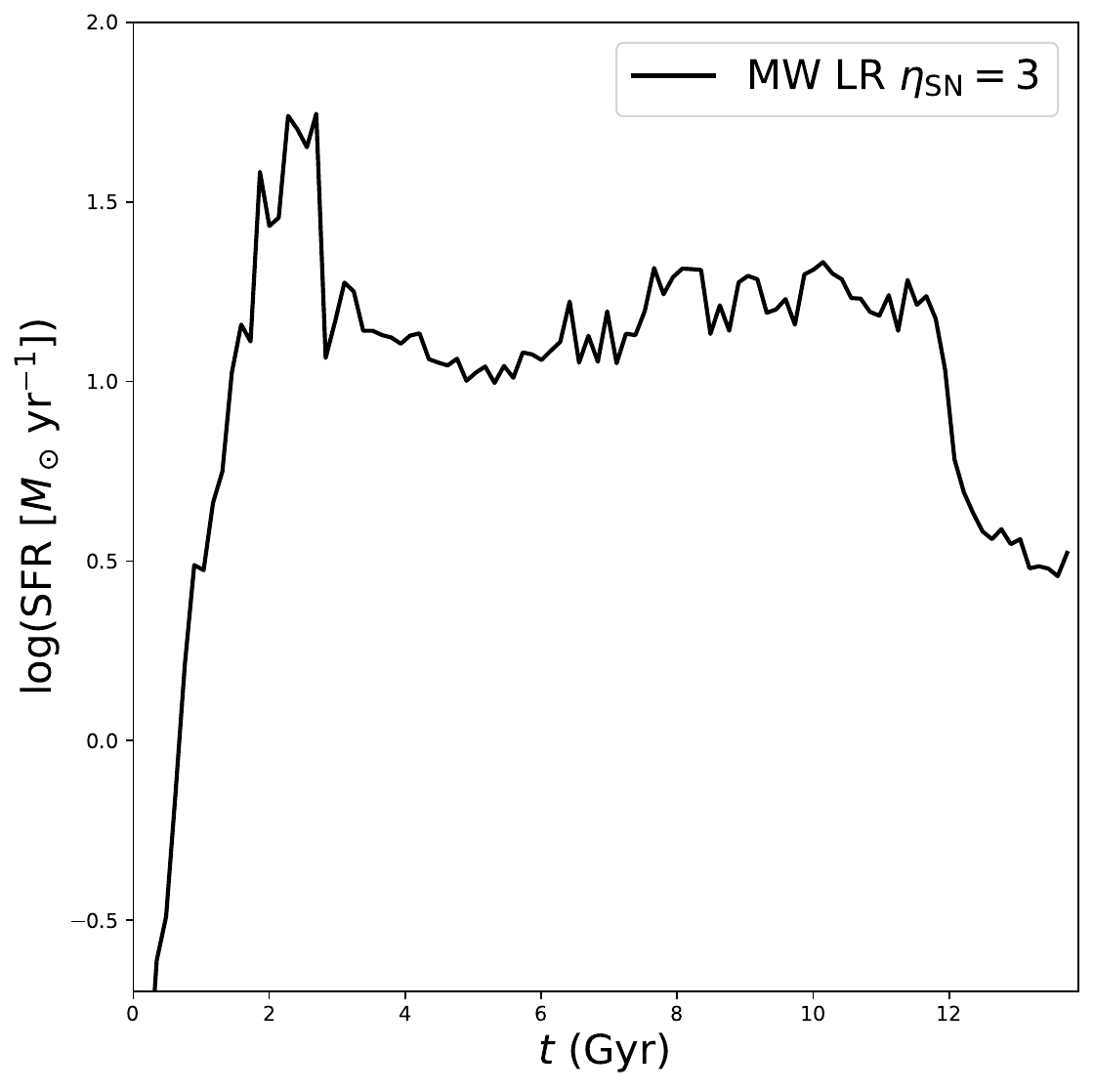} }
    \caption{Star formation history of the Milky Way-mass galaxy simulation with
    $\eta_{\mathrm{SN}}=3$ and LR resolution. The stars within the galactic radius at $z = 0$ are used. The vertical axis shows the star formation
    rate in logarithmic scale.
    \alttxt{Line plot of star formation rate history versus cosmic age for
a simulated Milky Way-mass galaxy.}
    \label{fig:sfh_mw}}
  \end{figure}
  In figure~\ref{fig:sfh_mw}, we present the star formation history (SFH) of the MW
  simulation. The SFR increases rapidly during the initial stages of galaxy
  assembly, reaching a peak of approximately $50\text{--}60\,M_{\odot}\,\mathrm{yr}
  ^{-1}$ at $t \sim 2.5\,\mathrm{Gyr}$. Following this peak, the galaxy maintains
  a relatively steady SFR of roughly $10\text{--}20\,M_{\odot}\,\mathrm{yr}^{-1}$
  for the majority of its evolution. In the final $2\,\mathrm{Gyr}$, the SFR exhibits
  a steep decline, reaching approximately $3\,M_{\odot}\,\mathrm{yr}^{-1}$ at
  $z = 0$ ($t \approx 13.8\,\mathrm{Gyr}$).

  This late-stage decline is associated with the cessation of star
  formation in the central region of the galaxy. At $t \simeq 10$~Gyr ($z \approx
  0.35$), a minor merger triggers a temporary enhancement in the SFR. However,
  the subsequent SN feedback, acting in concert with the earlier cumulative heating
  of the halo, effectively suppresses further gas accretion from the cosmic web.
  As the external gas supply is reduced, the gas reservoir in the disk is
  gradually exhausted. During this period, the gas disk relaxes from the
  distorted state induced by the merger into the settled, symmetric morphology
  at $z=0$ shown in figure~\ref{fig:map}. Due to the sustained high SFR over most of the
  Hubble time, the total stellar mass of this simulated galaxy is likely higher
  than that of the actual Milky Way.

\begin{figure}[h!]
    {\centering \includegraphics[width=8.5cm]{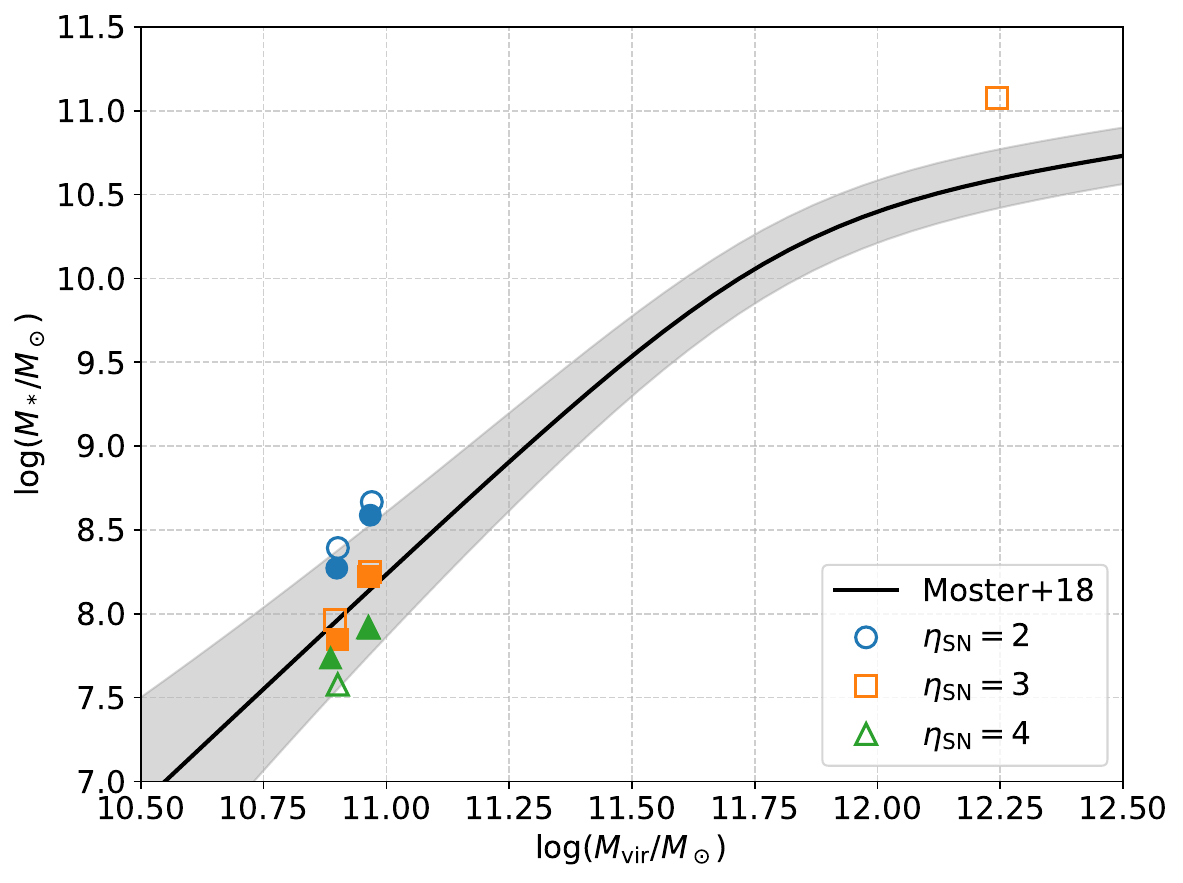} }
    \caption{Stellar mass--halo mass relation of the simulated galaxies at $z = 0.1$.
    The simulations with $\eta_\mathrm{SN} = 2$, $3$, and $4$ are indicated by circles, squares, and triangles, respectively. Open and filled symbols indicate LR and HR simulations, respectively.
    The stellar mass is defined as the mass of stars within the galactic radius.
    We also show the relation predicted by an empirical model at $z = 0.1$ \citep{moster2018}, with its 1\,$\sigma$ uncertainty represented by the shaded region.
    \alttxt{Scatter plot of stellar mass versus halo mass for simulated
galaxies compared to an empirical relation.}
    \label{fig:smhm}
    }
\end{figure}
  In figure~\ref{fig:smhm}, we show the stellar mass--halo mass (SMHM) relation of our simulated galaxies at $z = 0.1$. As expected from the sustained high star formation rates seen in figure~\ref{fig:sfh_mw}, the MW-mass simulation exhibits a stellar mass that is significantly higher than the median prediction of the empirical model by \citet{moster2018}.

Given that the feedback strength $\eta_{\mathrm{SN}} = 3$ reproduces the stellar masses of dwarf galaxies in our sample, the overproduction of stars in the MW-mass simulation suggests the necessity for additional feedback processes that preferentially operate in more massive systems. Indeed, numerous studies have indicated that feedback from active galactic nuclei (AGN) is essential at this mass scale to suppress star formation and match observed scaling relations (e.g. \cite{croton06, bower06, okamoto14, eagle, weinberger_2017_IllustrisTNG, nu2GCagn}).

\section{Discussion}
\label{Discussion}


This study presents a mechanical feedback scheme that accounts for the relative motion between gas and star particles, as well as the cumulative effect of multiple momentum injections. This scheme achieves excellent numerical convergence by preventing the unphysical over-injection of kinetic energy.


The simulated dwarf galaxies reproduce the stellar mass--halo mass relation predicted by the empirical model \citep{moster2018} with a feedback strength of $\eta_\mathrm{SN} = 3$.
However, the same model overproduces stars in the Milky Way-mass simulation.
This discrepancy is likely due to the absence of AGN feedback, which is known to play a crucial role at the $10^{12} M_{\odot}$ mass scale. While our SN feedback is powerful enough to create kpc-scale cavities and drive gas out of the interstellar medium (ISM), our analysis of the baryon fraction suggests it is unable to eject gas beyond the virial radius. Consequently, the expelled gas remains within the halo and eventually cools back onto the disk. This "gas recycling" sustains star formation rates of $10\text{--}20\,M_{\odot}\,\mathrm{yr}^{-1}$ for the majority of the galaxy's history. Indeed, numerous studies have indicated that feedback from AGN is essential at this mass scale to suppress star formation (e.g. \cite{croton06, bower06, okamoto14, dubois_2014_horizon_agn, eagle, weinberger_2017_IllustrisTNG, nu2GCagn}).

Interestingly, the star formation rate in the MW-mass simulation exhibits a steep decline at $t \approx 11.2$~Gyr ($z \approx 0.22$). This coincides with the exhaustion of the central gas reservoir and the morphological relaxation of the gas disk. Following a minor merger at $z \approx 0.35$ that temporarily distorted the disk and triggered a final burst of star formation, the galaxy settled into the symmetric, thin-disk morphology seen in figure~\ref{fig:map}. This transition suggests that while SN feedback regulates the "state" of the ISM, the overall quenching in massive systems is driven by the balance between gas consumption and the lack of fresh accretion from the halo.

Furthermore, the required SN feedback strength, $\eta_{\mathrm{SN}} = 3$, is significantly larger than the canonical value of unity. This discrepancy suggests that the high SN feedback strength may be compensating for missing physical processes. For example, our current model only crudely accounts for photoionization and radiation pressure. As noted by \citet{hu2017}, a more elaborate treatment of radiative feedback can enhance the suppression of star formation in dwarf galaxies. \add{Other missing processes include a top-heavy IMF in starbursts \citep{oka05}, which would increase the number of CCSNe per unit stellar mass, and the clustering of SN explosions within giant molecular clouds, which can boost the effective momentum per SN event \citep{gentry2017}. More elaborate treatment of early feedback processes \citep{benitez-llambay2026} would elevate the SN feedback efficiency by pre-processing the ISM, reducing the density into which SNe explode.}

Similarly, cosmic rays (CRs) have gained considerable attention for their ability to drive continuous, cold galactic outflows (e.g. \cite{jubelgas2008, pakmor2016, dashyan2020, hopkins2020}). Since CRs provide a non-thermal pressure support that does not suffer from rapid radiative cooling, they may be more effective at ejecting gas from the halo than thermal or mechanical SN feedback alone. While investigating the combined effects of CRs and our energy-conserving mechanical feedback is beyond the scope of this paper, our results underscore that a stringent treatment of SN energy injection is a prerequisite for accurately modeling these additional feedback channels.

\section{Conclusion}

We have developed a mechanical supernova feedback scheme that enforces stringent energy conservation. Our analysis reveals that without proper momentum correction, multiple SN events impacting a single gas element within a single timestep lead to an unphysical over-injection of kinetic energy, artificially inflating the feedback efficiency. We demonstrate that our proposed correction not only resolves this issue but also enhances numerical convergence. By summing the kinetic energy injections in the rest frame of the gas element, our method ensures exact energy conservation while avoiding the numerical pathologies associated with momentum inversion—a problem that can occur when calculating energy increments in the lab frame.

However, the momentum correction for multiple injections inherently leads to a violation of momentum conservation. When momentum is injected on a galactic scale without being strictly conserved, it can disturb the global angular momentum distribution and potentially hinder the formation of a stable gas disk. A common approach to mitigate this is the introduction of a maximum coupling radius (typically $\sim 1$~kpc). While this prevents momentum coupling with distant gas elements,
it can introduce numerical pathologies: as the number of coupled
gas elements within the maximum radius decreases, momentum
injection becomes increasingly anisotropic, and momentum
conservation cannot be maintained when only a single element
lies within the radius. In the most extreme case, no gas elements
exist within the maximum radius, and the feedback energy and
momentum must be handled through ad hoc procedures or are lost entirely.

To address this, we adopt a hybrid approach based on the resolution of the cooling radius. When the cooling radius is resolved by at least two
local inter-element separations, we employ pure thermal feedback instead of mechanical feedback. By switching to thermal injection in low-density environments—where the coupling radius would otherwise be unphysically large—we effectively limit the scale of momentum coupling. This transition also avoids the inappropriate application of mechanical feedback in the hot ISM, where equation~(\ref{eq:terminal}), derived from fits to simulations assuming a pressureless ambient medium, is no longer physically applicable.

Regarding numerical robustness, the star formation histories of dwarf galaxies simulated at two different mass resolutions (differing by a factor of 8) converge well with our new feedback scheme. However, we find that the spatial distributions of stars and gas within the virial radius differ notably; lower-resolution runs exhibit less centrally concentrated profiles compared to their high-resolution counterparts. To determine the resolution
required to produce reliable internal structures, more extensive convergence studies using even higher-resolution simulations are necessary.


\change{Our scheme is designed for high- to intermediate-resolution zoom-in
simulations with star particle masses up to $\sim 10^5\,M_\odot$.
This follows from the requirement that each star particle hosts at
most one SN event per timestep, which becomes increasingly
restrictive for larger star particle masses. In future work, we
plan to develop an improved scheme that relaxes the requirement of at most one SN event per star
particle per timestep to extend the
applicability to lower-resolution simulations and to speed up intermediate-resolution simulations.}

\section*{Funding}
This study is supported by JSPS/MEXT KAKENHI Grant Number JP25H00671
and by MEXT through the ``Program for Promoting Researches on the
Supercomputer Fugaku'' (Toward a unified view of the Universe: from
large-scale structures to planets, Grant No. JPMXP1020200109).

\begin{ack}
  We thank the anonymous referee for constructive comments 
that improved the manuscript.
  We also thank P. Hopkins for making GIZMO public.
  The numerical simulations were carried out on Cray XC50 and XD2000, and analyses were carried out on analysis servers at the Center for Computational Astrophysics, National Astronomical Observatory of Japan.
\end{ack}


\end{document}